\documentstyle[osa,manuscript,epsbox]{revtex}  

\begin{document}                

\title{Cellular Automaton Rule184++$C$.\\
A Simple Model for the Complex Dynamics of Various Particles Flow.}

\author{Akinori Awazu}

\address{Department of Physics \\
Ibaraki University, Mito 310-0056, Japan          } 

\maketitle
\begin{abstract}
A cellular automaton named Rule 184++$C$ is proposed as
a meta-model to investigate the flow of various complex particles.
In this model, unlike the granular pipe 
flow and the traffic flow, not only the free-jam phase transition but 
also the free-intermediate, the intermediate-jam, and the dilute-dense
phase transitions appear. Moreover, the freezing phenomena appear
if the system contains two types of different particles.
\end{abstract}

Recently, the flow of materials which consist of numerous discrete elements,
for example the granular pipe flow, 
the traffic flow, and so on are investigated analytically, experimentally,
and numerically
\cite{r1,r11,r3,r4,r5,r6,r7,r8,r9,r10,r12,y12}
. They succeeded to explain phenomena 
in these systems, e.g, the free-jam phase transition and the $1/f^{\alpha}$ 
fluctuation of the local density\cite{r3,r7,r12,r15}. However, it is fact that 
the behavior of the granular 
pipe flow depends on materials filled in the pipe\cite{r12,r15}.
For example, when the long range interactions like Coulomb force, 
the inhomogeneity of softness, or the form of materials is taken into
account, the system is expected to behave in more 
complex manner. Now, to discuss such complex flow of various
particles, we propose a simple meta-model which we name cellular 
automaton (CA) named Rule 184++$C$ . 

The dynamics of Rule 184++$C$ is based on that of Rule 184. The Rule 184   
is taken as one of the simplest models of the traffic and the granular flow.
Here, in addition to the Rule 184 dynamics, as the +$C$ rules, we
employ a set of simple rules for the velocity change of individual particles.
The dynamics of each particle is described by the equations;
\begin{equation}
v^{i}_{n+1} = F(v^{i}_{n}, v^{i+1}_{n}, d^{i}_{n})
\end{equation}
\begin{equation}
x^{i}_{n+1} = x^{i}_{n} + v^{i}_{n+1}
\end{equation}   
where we number the particles $i$ from the upper part of the traffic
stream to the downward. 
The quantities $x^{i}_{n}$ and $v^{i}_{n}$ are the position   
and the velocity of the $i$th particle at time step $n$, and
$d^{i}_{n}$ is the number of empty sites between $i$th and $i+1$th
particle.  
The function $F(v^{i}_{n}, v^{i+1}_{n}, d^{i}_{n})$ and the velocity
$v^{i}_{n}$take 0 or 1, whereas $F(*)$ obeys the following rules;

I)When $d^{i}_{n}>1$, $F(v^{i}_{n}, v^{i+1}_{n}, d^{i}_{n})=1$ always holds, 
and when $d^{i}_{n}=0$, $F(v^{i}_{n}, v^{i+1}_{n}, 0)=0$ always holds.

II)When $d^{i}_{n}=1$, $F(v^{i}_{n}, v^{i+1}_{n}, d^{i}_{n}=1)$ takes 0 or 1 which 
depends on the type of the particle at $i$.
Since $v^{i}_{n}$ takes 0 or 1, the combination of $(v^{i}_{n},
v^{i+1}_{n})$ takes one of the following four patterns, $(0, 0)$, $(0,
1)$, $(1, 0)$, $(1, 1)$. For each combination , $F$ takes the value 0
or 1. Hence, 16 types of rules for  
$F(v^{i}_{n}, v^{i+1}_{n}, 1)$  (from \{ 
$F(0, 0, 1)=0$, $F(0, 1, 1)=0$, $F(1, 0, 1)=0$, $F(1, 1, 1)=0$ \} to 
\{ $F(0, 0, 1)=1$, $F(0, 1, 1)=1$, $F(1, 0, 1)=1$, $F(1, 1, 1)=1$
\})are considerd. 
In other words, we define 16 types of particles according to 
the function $F(v^{i}_{n}, v^{i+1}_{n}, 1)$. 
Now, we name the types of particles using
the rule number $C$ which is defined like Wolfram's method \cite{r0}, 
\begin{equation}
C=2^{0}F_{C}(0, 0, 1)+2^{1}F_{C}(0, 1, 1)+2^{2}F_{C}(1, 0, 1)+2^{3}F_{C}(1, 1, 1)
\end{equation}
If the type of all particles is $C=15$, the dynamics is same as CA Rule 184. 
The boundary condition is set periodic and the positions and the velocities 
of particles are set random at initial conditions. Hereafter, for the
first, simulation results for pure systems in which all particles
have same $C$ are introduced. Also the statistical and the dynamical
properties for each stationary state are discussed.  
Second, the simulation of mixed systems in which various $C$  
particles coexist are discussed.

In Figure 1 are the typical fundamental diagrams which are the 
relations between the particle density of the system $\rho$ and the flow $f$ for each $C$. Here, $\rho$ is defined as 
(the number of the occupied sites by particles)/(the total number of
sites), and $f$ is defined as $<\sum_{i}v^{i}_{n}>$/(the total number of sites)
where $<..>$ means the time average. These 
fundamental diagrams are classified into following three types; (i)the
two phases type (2P-type) where $C=0,4,5,6,8,10,12,13,14,$ or $15$,
(ii)the three phases type (3P-type) where $C=2,9,$ or $11$, and 
(iii)the four phases type (4P-type) where $C=1,3,$ or $7$. The property of
the steady state for the
2P-type is following; When $\rho$ is low, all
particles move at each time step, while slugs appear in the system if
$\rho$ is higher than a critical value. Here, the term 'slug' means an 
array of particles which do not move i.e., $v^{j}_{n}=0$
$v^{j+1}_{n}=0 \dots$ . Generally, slugs move backward in the traffic
stream keeping the spatial pattern of them invariant. Each of them is
kept by the balance between the incoming free-flow particles from the
upward and the outgoing free-flow particles to the downward.
As such, with $\rho$ increases, the phases transition from
the 'free-flow state' (without slug) to the 'jam-flow state' (with
slugs) occurs. This property is qualitatively same as known results
of recent traffic and granular flow
models\cite{r1,r11,r3,r4,r5,r6,r7,r8,r9,r10,r12}.
In particular, the dynamics of the pure particle systems with $C=12$
($F_{12}(0, 0, 1)=0$, $F_{12}(0, 1, 1)=0$, $F_{12}(1, 0, 1)=1$,
$F_{12}(1, 1, 1)=1$)is equivalent to the dynamics of the
deterministic traffic flow model proposed by Takayasu and Takayasu\cite{r7}. 
Different from such systems, 
the 3P-type systems include the parameter region of the
'intermediate-flow state'. 
At the intermediate $\rho$ values between those
of the free-flow state and of the jam-flow state, the third state
which is different from the former two states, takes place as shown in
Fig.2 (b) and (e).
Among the 3P-type systems, we discuss the pure systems with
$C=11$ and $C=9$.
Figure 2 (a) (b) and (c) show the space-time evolutions of the 
stationary states of the pure systems with $C=11$ ($F_{11}(0, 0,
1)=1$, $F_{11}(0, 1, 1)=1$, $F_{11}(1, 0, 1)=0$, $F_{11}(1, 1, 1)=1$).
They respectively correspond to (a)the free-flow state, (b)the
intermediate-flow state and (c)the jam-flow state. 
Here, dots represent individual particles where black dots indicate
$v=0$ particles and gray dots are $v=1$
particles. The behaviors of particles in (a) and (c) of Fig.2 are qualitatively
same as the space-time evolutions of the 2P-type systems.
In the free-flow state, more than two successive empty sites appear in 
front of all particles because
the particle density is low.
In the jam-flow state, slugs appear and survive stably
because the particle density is high.
Different from these two kinds of states, in the intermediated-flow state,
unstable slugs and more than two successive empty sites coexist (Fig.2 (b)).
Figure 2 (d) (e) and (f) show the space-time evolutions of the 
stationary states of the system with the $C=9$ particles ($F_{9}(0, 0, 1)=1$,
$F_{9}(0, 1, 1)=0$, $F_{9}(1, 0, 1)=0$, $F_{9}(1, 1, 1)=1$). There are 
two types of slugs. Isolated slugs consist of only one
particle $j$ with $v^{j}_{n}=0$, whereas large slugs consist of more than
two particles with $v^{j}_{n}=0$, $v^{j+1}_{n}=0 \dots$ .
Moreover the backward propagation velocity of large slugs 
is slower than that of isolated ones.
In the jam-flow state(Fig.2(f)), some large slugs remain stable in the system.
In the intermediate-flow state, however, large slugs are unstable and
repeat creation and annihilation irregularly in space and time (Fig.2 (e)). 
 
Next, we discuss the properties of the 4P-type systems. As an
example, the system with $C=3$  particles ($F_{3}(0, 0, 1)=1$, $F_{3}(0, 1,
1)=1$, $F_{3}(1, 0, 1)=0$, $F_{3}(1, 1, 1)=0$) is considered.  In figure 3
are typical space-time evolutions  
for several typical densities. When density $\rho$ is low ($0 < \rho
<\frac{1}{3}$), the free-flow state is 
realized, and when $\rho$ increases ($\frac{1}{3} < \rho <\frac{2}{5}$)
some slugs emerge, which means the free-jam transition takes place like in 
the 2P-type systems (Fig.3 (a),(b)). Within slugs of this
$\rho$ region the gap $d^{j}_{n}$ between particle $j$ and $j+1$ is
not zero, but repeats $d^{j}_{n}=1$ and $d^{j}_{n+1}=2$ by
turn. Moreover, the spatial pattern in this slug is periodic with the
unit $v^{j}_{n}=1$, $v^{j+1}_{n}=0$, $d^{j}_{n}=1$ and $d^{j+1}_{n}=2$.
In other words, these slugs are more dilute than what we see in the
2P-type and the 3P-type systems wherein gaps $d$
are all 0. Thus, we name these slugs 'dilute slugs' and this
state the 'dilute jam-flow state'.
When $\rho =\frac{2}{5}$, the
system is completely filled with dilute slugs. If $\rho$ increases
more ($\frac{2}{5} < \rho <\frac{2}{3}$), unlike the 2P-type
systems, different type of slugs from the dilute slugs appear (Fig.3
(c)). In these slugs, the spatial pattern is
periodic with the unit $v^{j}_{n}=1$ and $v^{j+1}_{n}=0$ similarly in
the dilute slug. However, in this case, the gaps in front of particles $j$ and
$j+1$ repeat ($d^{j}_{n}=0$, $d^{j+1}_{n}=1$) and ($d^{j}_{n+1}=1$,
$d^{j+1}_{n+1}=0$) by turn.
Moreover, the direction of movement of these slugs is downward, and
these slugs are surrounded by dilute slugs. Now, we
name these unusual slugs 'advancing slugs' and this state the 'advancing 
jam-flow state'.
Moreover, in the advancing jam-flow state, the flow increases in
proportional to $\rho$, in which sense the advancing jam-flow state is
similar to the free-flow state. 
When the density increases more ($\frac{2}{3} < \rho $), slugs in which 
the gaps $d^{i}_{n}$ are zero appear (Fig.3(d)),  
and the flow turns again to a decrease function of $\rho$. We call
this state the 'hard jam-flow state'.
As such, with the increase of $\rho$, three phase transitions, the free -
dilute jam, the dilute jam - advancing jam, and the advancing jam - hard jam
transitions appear. The characters of the first and the third are like 
what we see in the 2P-type system because flow sharply changes from
the increasing function of $\rho$ to decreasing function at the
transition point. So both of them are, in a wide sense, the free-jam 
transitions.
On the other hand, the second is the transition between the low density
(dilute) regime and the high density (dense) regime.
The similar transitions also appears for $C=1$ and $C=7$ systems.

It is noted the 3P-type systems and the 4P-type systems, respectively, have
common rules which hold throughout each of them.
The 3P-type systems share the rules
$F_{C}(0, 0, 1)=F_{C}(1, 1, 1)$ and $F_{C}(1, 0, 1)=0$. These rules
mean that when a particle comes close to the preceding particle,
repulsive force works between the two.
On the other hand, the 4P-type systems share the
rules,  $F_{C}(0, 0, 1)=1$ and $F_{C}(1, 1, 1)=0$.
Here, the rule $F_{C}(0, 0, 1)=1$ indicates that effective attractive
force works to a particle when this particle and the preceding one are
close each other and they are at a standstill
(i.e. $v^{i}_{n}=v^{i+1}_{n}=0$).  Similarly, $F_{C}(1, 1,
1)=0$ indicates that effective resistance acts on the rear particle
when this particle and the preceding one are
moving.(i.e. $v^{i}_{n}=v^{i+1}_{n}=1$)
According to such effective forces, above mentioned several flow
phases are realized.

So far we have discussed the statistical aspects of our system. In the 
next, we consider the dynamical properties of the above systems. For this
purpose, apart from the classification of the system according to 
the fundamental diagrams, the space-time evolutions are divided into three 
types. i) The regular flow regime ($C=0,2,3,5,8,10,11,12,14,$ or $15$): where
space-time evolution is regular and the flow $f$ is constant with time
(As shown in Fig. 2 (a) (b) (c), and Fig. 3).  ii) The oscillatory
flow regime ($C=1,4,7,$ or $13$):
where $f$ oscillates with time. For example in the $C=7$ particle
system, the velocities of individual particles in dilute
slugs synchronize each other to oscillate between 0 and 1 (Fig. 4
(a)). Therefore, $f$ oscillates with large amplitude (Fig. 4 (b)). iii)
The chaotic flow regime ($C=6,$ or $9$):  
The space time evolution of particles and the time evolution of $f$
are chaotic. (Examples are shown in Fig. 2 (e), (f), and Fig. 5(a).)
In particular, in the $C=9$ particles system, the
flow has $1/f$ fluctuation near the critical density ($\rho \sim
0.38$) of the phase transition between the intermediate-flow state and the
jam-flow state (Fig.5(b)).
Both type of particles (i.e. $C=6$
and $C=9$) share the symmetric dynamics
$F_{C}(0, 0, 1)=F_{C}(1, 1, 1) \ne F_{C}(0, 1, 1)=F_{C}(1, 0, 1)$.
In other words, only such systems that contain particles
with symmetric rules behave chaotic if the system is pure.

Finally, we focus 
the mixed systems in which two values of $C$ particles are mixed and
compare the behavior of them to those of 
the pure systems. Figure 6 are the typical fundamental 
diagrams of two types of pure systems and that of mixed systems of
these two types of particles. Here, the ratio of two types of
particles is 1:1. Almost all the cases, the relations
like Fig.6 (a) are realized. However, the relations like Fig.6 (b) also
are realized for some cases. Here, the flow of the mixed systems is
smaller than that of respective pure particles systems.
To discuss them, as an example, we
consider mixed systems with $C=2$ and $C=4$ particles (For $C=2$ :
$F_{2}(0, 0, 1)=0$, $F_{2}(0, 1, 1)=1$, $F_{2}(1, 0, 1)=0$, $F_{2}(1,
1, 1)=0$. For $C=4$ : $F_{4}(0, 0, 1)=0$, $F_{4}(0, 1, 1)=0$, $F_{4}(1, 0, 1)=1$, $F_{4}(1, 1, 1)=0$).  
Figure 7 (a) is a fundamental diagram of the almost pure $C=2$ particles
system except one $C=4$ particle inside it. Compared with the pure $C=2$
particles system, the
flow is drastically little, and in particular, no flow for $\rho \ge
0.5$. The origin of such behavior is following. When $\rho$ is
sufficiently large, that is, most of the 
gaps $d^{i}_{n}$ between successive particles are not larger than 1,
the dynamics of each particle mainly  
obeys the function $F_{C}(v_{n}^{i}, v_{n}^{i+1}, 1)$.
Once the $C=4$ particle stops, this particle remains stationary if
$d^{i}_{n}$ remain not larger than 1. 
Moreover, in such $d^{i}_{n}$, $C=2$ particle remains stationary when 
the preceding particle does not move.
In other words, a $C=4$ particle works as the coagulant of $C=2$
particles, and change the characters of the whole system.
Such relation appears also in different pairs of
particles e.g., $C=6$ and $C=4$ for $\rho>0.66$ (Fig.7 (b)), and $C=2$
and $C=12$. The mechanism of the freezing phenomena in these systems is
qualitatively same as the above. Moreover, $C=4$ and $C=12$ particles,
both of which work as the coagulant in the above mixed systems, share the
common rules $F_{C}(0, 0, 1)$=0 and $F_{C}(0, 1, 1)=0$.
     
To summarize, various properties of complex materials flow are 
realized using a meta-model named CA Rule 184++$C$.
Unlike the previous CA models of the granular particles flow or the
traffic flow, two types of novel
relations between the density and the flow are realized in some
types of particles systems. One is
realized in the three phases type systems in which the free-intermediate and
the intermediate-jam phase transitions occur with the increase of particle
density. The other is observed in the four phases type systems in which
not only the free-jam but also the dilute-dense transitions are
realized.
When these systems are 
classified from another point of view, there are three types of flow
regimes; the regular 
flow regime, the oscillatory flow regime, and the chaotic flow
regime. The particles in the 
chaotic flow systems share the symmetric rules. The causal relationship between
such symmetric rule and the chaotic behavior remains to be studied in
future.
Moreover, when two types of
particles are mixed in a system, one may works as a coagulant of the
other. In other words, the characters of pure particles systems are
drastically changed by the addition of only a few number of different type of particles.

The author is grateful to H.Nishimori for useful discussions. This
research was supported in part by the Ibaraki University SVBL.

\newpage

Fig.1: Typical fundamental diagrams for each $C$. There are three types 
of fundamental diagrams, (a)(b)(c) 2P-type, (d)(e)(f) 3P-type, 
and (g)(h)(i) 4P-type.
\vspace{8mm}

Fig.2: Space-time evolutions of the stationary states of
$C=11$ particles systems ((a),(b),(c)), and $C=9$ ((d),(e),(f)), 
where black dots means $v=0$ particles and gray dots means $v=1$ 
particles. They indicates, respectively, the free-flow ((a) and (d)),
the intermediate-flow ((b) and (e)), and the jam-flow ((c) and (f)).   
\vspace{8mm}

Fig.3: The space-time evolutions of the stationary states of $C=3$ particles 
systems, respectively, (a) the free-flow, (b) the dilute jam-flow,
(c)the advancing jam-flow, 
and (c)the hard jam-flow.  
\vspace{8mm}

Fig.4: (a) Space-time evolutions of the stationary states of $C=7$
particles systems. (b) Fundamental diagram of $C=7$ particles system
at even time and odd time. 
\vspace{8mm}

Fig.5: (a) Fundamental diagram of $C=9$ particles system.
(b) Power spectrum of the flow fluctuation of $C=9$ 
particles system near the critical density ($\rho\sim 0.38$) of
intermediate-jam transition. 
\vspace{8mm}

Fig.6: Typical fundamental diagrams of respective two pure particles
systems and the mixed particles systems. The ratio of particles is
1:1. (a) Normal type. (b)Decreasing (by the mixing) type.
\vspace{8mm}

Fig.7: Fundamental diagrams of the 'almost pure' systems; (a) $C=2$
particles system and the $C=6$ particles system with one $C=4$
particle inside of each. The doted line across the $\rho$ axis with
$\rho=0.66$.

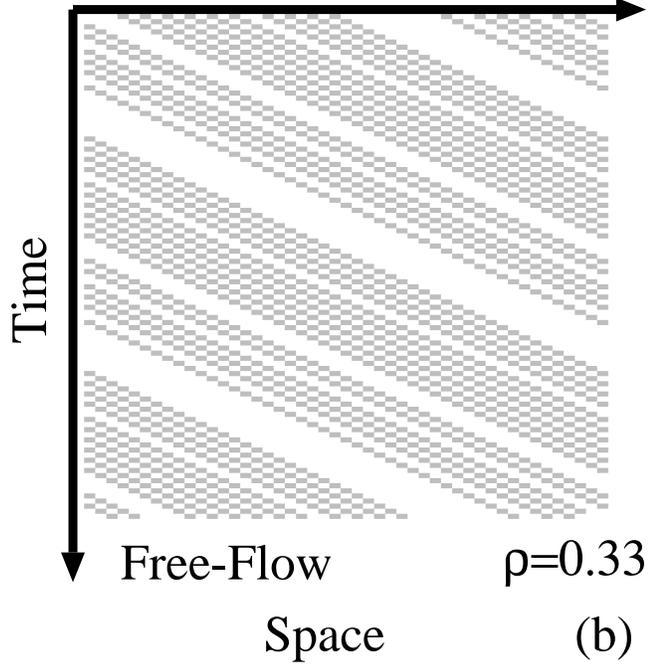
\end{document}